\documentclass[conference]{IEEEtran}
\IEEEoverridecommandlockouts
\usepackage[utf8]{inputenc}
\usepackage{cite}
\usepackage{amsmath,amssymb,amsfonts}
\usepackage{algorithmic}
\usepackage{graphicx}
\usepackage{float}
\usepackage{textcomp}
\usepackage{xcolor}
\usepackage{booktabs}
\usepackage{tabularx}
\title{LSTM Framework for Classification of Radar and Communications Signals \\ \thanks{This work was supported by project PID2020-113979RB-C21 founded by MCIN/AEI/10.13039/501100011033.}}

\author{
    \IEEEauthorblockN{
        Victoria Clerico\IEEEauthorrefmark{1} (Student Member, IEEE),
        Jorge González-López\IEEEauthorrefmark{1},\\
        Gady Agam\IEEEauthorrefmark{2},
        Jesús Grajal\IEEEauthorrefmark{1} (Senior Member, IEEE)
    }
    \\
    \IEEEauthorblockA{
        \IEEEauthorrefmark{1}Information Processing and Telecommunications Center, Universidad Politécnica de Madrid.  \\E.T.S.I. Telecomunicación, Av. Complutense 30, 28040 Madrid, Spain.\\
        Corresponding author: Victoria Clerico (e-mail: mclerico@ieee.org)
    }
    \IEEEauthorblockA{
          \IEEEauthorrefmark{2}Department of Computer Science, IIT, Chicago, USA
    }  
}

\date{November 2022}

\begin{document}
\maketitle
\begin{abstract}
Although radar and communications signal classification are usually treated separately, they share similar characteristics, and methods applied in one domain can be potentially applied in the other. We propose a simple and unified scheme for the classification of radar and communications signals using Long Short-Term Memory (LSTM) neural networks. This proposal 
provides an improvement of the state of the art on radar signals where LSTM models are starting to be applied within schemes of higher complexity. To date, there is no standard public dataset for radar signals. Therefore, we propose DeepRadar2022\footnotemark[1], a radar dataset used in our systematic evaluations that is available publicly and will facilitate a standard comparison between methods.
\footnotetext[1]{Available for download in https://www.kaggle.com/datasets/khilian/deepradar}
\end{abstract}


\section{Introduction}

Automatic modulation classification (AMC) consists in automatic determination of the modulation of a series of collected samples. It is the step that follows the detection of the signal and that is needed for data demodulation, therefore, it plays an important role in many civilian and military receivers \cite{b0}.

Taking into account the classical approaches, AMC algorithms can be classified into two categories: those based on the likelihood function (LB, `Likelihood-Based') and those based on feature extraction (FB, `Feature Based') \cite{b1}.
The first offers the optimal solution by minimizing the probability of false classification through the assumption that the probability density function (PDF) contains all the information needed for a specific waveform. Therefore, classification is performed by comparing the PDF likelihood ratio with a decision threshold \cite{b1}. The problem lies
in its high computational complexity, which means that this method could not be suitable for real working environments.
In contrast, FB algorithms extract representative features of each type of signal for their subsequent classification. These algorithms are
suboptimal but are often preferred because they are easy to implement and suitable for
real-time applications \cite{b1}.
Despite that, FB classifiers rely heavily on expert knowledge, so even though they are a good approximation on specific environments, they are highly complex and require a lot of time for development. 
While these algorithms have been successfully implemented to develop AMCs, Machine Learning (ML) and Deep Learning (DL) are considered good alternatives to develop high-performance and accurate AMCs without the need of time-consuming classical approaches. Algorithms such as K-Nearest Neighbors \cite{b3},  Support Vector Machine \cite{b3,b4}, Multilayer Perceptron (MLP) \cite{b5}, Recurrent Neural Networks (RNN) \cite{b6,b44} and Convolutional Neural Networks (CNN)\cite{b7,b8,b9,b10} have recently been used for this purpose.

Previous literature shows the success of LSTM networks in processing and classifying sequences. Thus, our objective is to provide a robust and simplified AMC based on LSTM networks for both communications and radar signals. To do so, the public RadioML 2018.01A communications dataset \cite{b8} was used. Since there is no public radar signal dataset, we have created and published DeepRadar2022\footnotemark[1], a radar dataset with continuous and pulsed signals of 23 classes. Furthermore, for comparison purposes, a dataset of eight types of radar signals was reproduced, which was proposed in the current state-of-the-art literature on radar signal classification \cite{b44}.

The remainder of the paper is distributed as follows. The most relevant proposals and studies on AMCs with DL and ML are outlined in Section II. The signal model and datasets are introduced in Section III. Afterwards, our neural network architecture, metrics, and experimentation are described in Section IV, and the results are presented in Section V. Finally, the main conclusions of this work are presented in Section VI.

\hfill
\section{Related Work}
Despite the usefulness of the LB and FB classification algorithms, the appearance of artificial intelligence has revolutionized many areas of interest. ML and DL tools have been used in the past to build modulation classifiers for communications and radar signals but have been treated separately.

When classifying communications signals, the most typical signal representation is time series, and the most widely used method for its classification has been one-dimensional CNN \cite{b7,b11, b27,b28,b29,b30,b31,b32,b33,b34,b35}. Similarly, recent proposals such as GGCNN \cite{b40} and SE-MSFN \cite{b41} used RadioML 2018.01A providing outperforming results.  

In contrast, radar signals have been classified by feature extraction and some MLP architecture \cite{b5} or by using recorded signals as time-frequency images to be used in conjunction with two-dimensional CNN \cite{b3,b4,b5, b15} or mixed CNN with tree structure-based machine learning process optimization classifier \cite{b23}, CNN with Reinforcement Learning \cite{b24}, CNN with denoising autoencoders \cite{b10, b25, b26} and CNN with LSTM \cite{b21,b22}. Some examples of pre-trained models based on CNNs that have been used lately for radar classification are AlexNet \cite{b9} and Inception \cite{b10}. In addition to that, certain attempts have been made to use RNN in recent years by processing signals as time series, some of them using LSTM with attention mechanisms \cite{b44} or gated recurrent unit networks \cite{b16,b17,b18}.

\section{Signal model and datasets}
To perform our experiments, one or more data sources should be chosen with a large number of signals per class. 
\subsection{Signal Model}
The datasets are composed of radar or communication signals. These signals are defined by the following model:

\begin{equation}
s(t) =  A s_n(t) + r(t) 
\end{equation}

\noindent
where $s_n(t)$ represents a normalized signal of unit power,  $A$ corresponds to a scale factor of the signal power, and $r(t)$ is Complex Additive Gaussian Noise (CAWN) with real and imaginary parts with variance $\sigma^2$. Consequently, the resulting SNR is established as follows.

\begin{equation}
snr = \frac{A^2}{2\sigma^2}
\end{equation}

All signals have a 1024x2 size corresponding to the In-Phase and Quadrature of the time samples. Data bases of equal-sized signals are required to process the sequences through our LSTM framework, which has a fixed input size. Hence, the three datasets used for our experiments are presented below. 

\subsection{Datasets}
\begin{itemize}
    \item Communications signals: These signals come from the RadioML 2018.01A dataset\footnote{Available for download in www.deepsig.ai/datasets}. This dataset contains 2555904 signals with an SNR range between -20 dB and 30 dB with a 2 dB step. There are 24 different digital and analog modulations: OOK, 4-8 ASK, 2-32 PSK, 16-128 APSK, 16 - 256 QAM, AM-SSB-WC, AM-SSB-SC, AM-DSB-WC, AM-DSB-SC, FM, GMSK and OQPSK. In addition to signals with synthetic channel effects generated with GNU Radio, it also includes over-the-air (OTA) recordings \cite{b8}. Its main limitation is that neither the parameters regarding the generation of the signals nor the origin of each signal, a synthetic recording or an OTA capture, were specified.
    
    \item Radar signals: We created \textbf{DeepRadar2022}, a dataset that contains time sequences of radar signals considering the 23 modulations given in Table \ref{tab1} with all their parameters listed: sampling frequency ($f_s$), carrier frequency ($f_c$), pulse-width ($PW$), bandwidth ($BW$), symbol rate ($v_s$), length of Barker ($L_c$), number of carrier cycles in a single-phase symbol ($M$), relative prime of M ($r$), amplitude of secondary lobes ($S$), number of segments ($N_g$), phase states ($PS$) and frequency variation ($\Delta f$). Note that $f_s$ and the input sequence are fixed to 100 MHz and 1024 samples, respectively. However, modulations with few symbols and high symbol rates lead to short signals in time and, therefore, to a small number of samples. Thus, we perform an interpolation to obtain a set of equal-sized signals. Finally, we have created 469200 signals for the training set and 156400 signals for the validation and testing set each. As a result, the entire dataset contains 782000 signals. 
\end{itemize}
\begin{table}[htbp]
	\caption{Summary of the DeepRadar2022 dataset modulations and their respective parameters.}
	\begin{center}
		\begin{tabular}{|c|c|c|c|}
			\hline
			Modulation type & Parameters & Values \\ \hline
			& \begin{tabular}[c]{@{}l@{}}$f_s$\\ input size\\ SNR\\ $f_c$ \end{tabular} & \begin{tabular}[c]{@{}l@{}}100 MHz\\  1024 samples\\-12:2:20 dB\\ U(-$f_s$/4, $f_s$/4)\end{tabular} \\ \hline
			NM & - & - \\ \hline
			LFM & \begin{tabular}[c]{@{}l@{}}BW\\    \end{tabular} & \begin{tabular}[c]{@{}l@{}}U($f_s$/20, $f_s$/4)\\    \end{tabular} \\ \hline
			PSK & \begin{tabular}[c]{@{}l@{}} Order\\ $v_s$\\ \end{tabular} & \begin{tabular}[c]{@{}l@{}}\{2,4,8\}\\ \{2, 5, 10, 15, 20\} \\ \end{tabular} \\ \hline
			Barker & \begin{tabular}[c]{@{}l@{}} $L_c$  \\ $v_s$\end{tabular} & \begin{tabular}[c]{@{}l@{}}\{5,7,11,13\}  \\ \{2, 5, 10, 15, 20\} \end{tabular} \\ \hline
			Frank, P1, Px & \begin{tabular}[c]{@{}l@{}} M  \\ $v_s$\end{tabular} & \begin{tabular}[c]{@{}l@{}}\{4, 5,   6, 7, 8\}  \\ \{7, 10,   15, 20\} \end{tabular} \\ \hline
			P2 & \begin{tabular}[c]{@{}l@{}} M \\ $v_s$\end{tabular} & \begin{tabular}[c]{@{}l@{}}\{4, 6, 8\}    \\ \{7, 10, 15, 20\} \end{tabular} \\ \hline
			P3, P4 & \begin{tabular}[c]{@{}l@{}}M   \\ $v_s$ \end{tabular} & \begin{tabular}[c]{@{}l@{}}\{16, 25, 36, 49, 64\}  \\ \{7, 10,   15, 20\} \end{tabular} \\ \hline
			Zadoff-Chu & \begin{tabular}[c]{@{}l@{}}M\\ r \\ $v_s$\end{tabular} & \begin{tabular}[c]{@{}l@{}}\{16, 25, 36, 49, 64\}  \\ \{11, 13\}\\ \{7, 10, 15, 20\} \end{tabular} \\ \hline
			Huffman & \begin{tabular}[c]{@{}l@{}}M\\ $v_s$  \\ S \end{tabular} & \begin{tabular}[c]{@{}l@{}}\{16, 25,   36, 49, 64\}  \\ \{7, 10,   15, 20\} \\ \{-63,-60,-56\}   dB\end{tabular} \\ \hline
			T1-T2 & \begin{tabular}[c]{@{}l@{}}$N_{g}$ \\ PW   \\PS\end{tabular} & \begin{tabular}[c]{@{}l@{}}\{4, 5, 6\}   \\ U(256, 1024) samples\\ 2\end{tabular} \\ \hline
			T3-T4 & \begin{tabular}[c]{@{}l@{}}BW  \\ PW  \\ PS \end{tabular} & \begin{tabular}[c]{@{}l@{}}U($f_s$/20, $f_s$/4)    \\ U(256,1024) samples\\ 2\end{tabular} \\ \hline
			FSK & \begin{tabular}[c]{@{}l@{}}Order\\ $v_s$   \\ $\Delta f$\end{tabular} & \begin{tabular}[c]{@{}l@{}}\{2,4,8\}    \\ \{1, 2, 5, 10, 15\}    \\ $v_s$  \end{tabular} \\ \hline
			Costas & \begin{tabular}[c]{@{}l@{}} M   \\ $v_s$   \\ $\Delta f$\end{tabular} & \begin{tabular}[c]{@{}l@{}}\{3,4,5,6\}   \\ \{1, 2, 5, 10, 15\}   \\ $v_s$ \end{tabular} \\ \hline
			Noise (CAWN) & $\sigma$ & 1 \\ \hline
		\end{tabular}
  \label{tab1}
	\end{center}

\begin{minipage}{3.5in}
\vspace{1mm}
$v_s$ values are presented in Msymb/s. $M$ values are presented in symbols. U(.) stands for a random uniform distribution.
\end{minipage}

\end{table}

Furthermore, we recreated the 8-class signal dataset proposed by the authors of SABLNet \cite{b44}. These signals have SNRs between -20 and 20 dB with 2 dB step. Following their specifications, the resulting signals are generated with CW, LFM, BFSK, SIN, EXP, SFW, BPSK and BASK modulations. In addition, the authors proposed the classification of the signals in two input formats: raw time sequences and after a preprocessing step with the autocorrelation computation.

\section{Long Short-Term Memory Framework}

\begin{figure}[h!]
	\centering
	\includegraphics[width=0.45\textwidth]{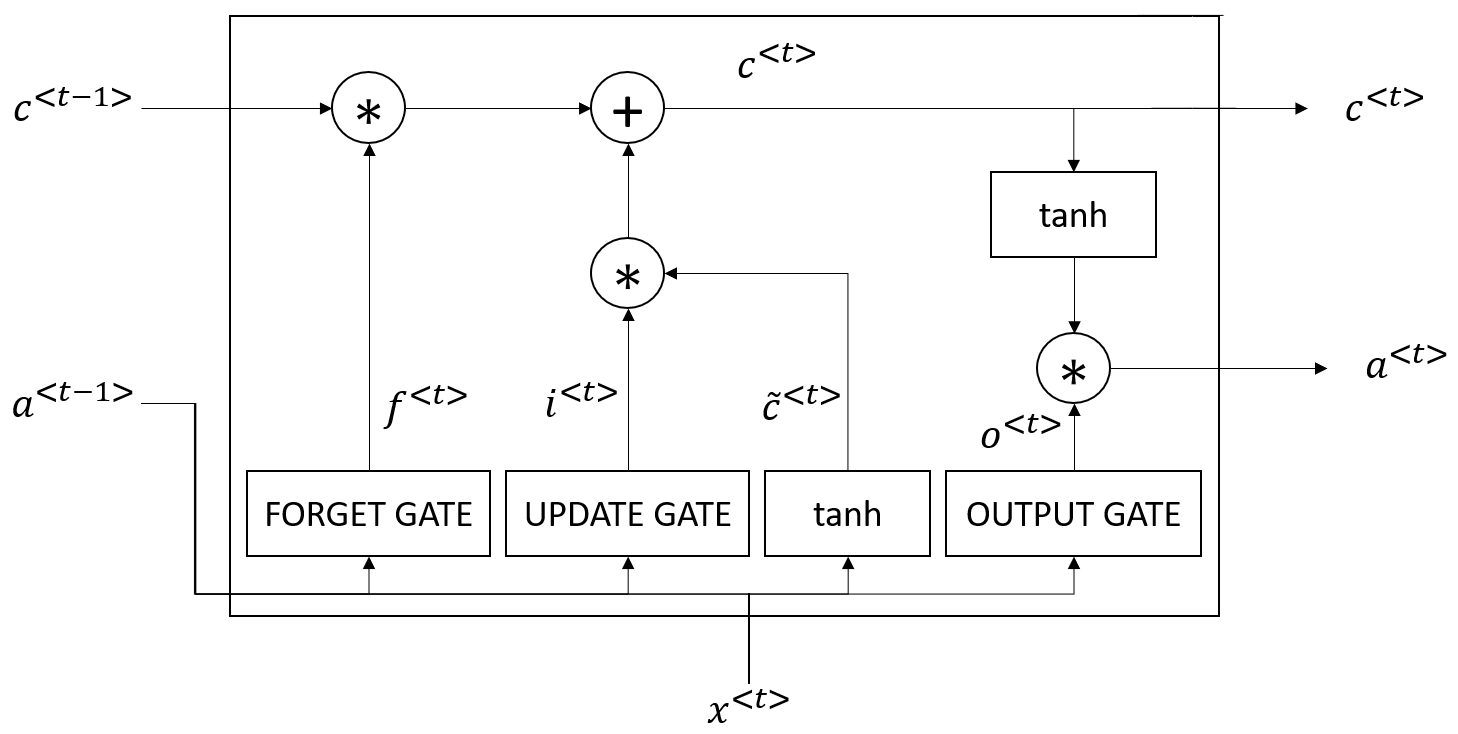}
	\caption{Internal functioning of a LSTM cell. The temporal parameter t is defined to represent the input element of the temporal series.}
	\label{fig:lstm}
\end{figure}

In recent years, the use of RNNs has been introduced. RNNs are derived from an attempt to correct the lack of memory of convolutional networks. Instead of processing the input data as a whole, they iterate through the sequence elements, keeping the information regarding all the previous elements. However, the major problem with recurrent neural networks is the vanishing gradient, since the input sequences can be very long.
LSTM networks solve the vanishing gradient issue, which is particularly troublesome. They consist of four different gate units, Figure \ref{fig:lstm}, and two different memory states, long-term ( memory cells, $c^{<t>}$) and short-term (activations, $a^{<t>}$). This way, they deal with the vanishing gradient by discarding the non-valuable information while keeping the important one \cite{bb4}. This aspect is what makes LSTM networks a very suitable approach to our classification problem when processing signals as sampled time sequences.

Figure \ref{fig:lstm} represents the internal functioning of each of the cells that make up the LSTM layer. These cells are formed by a more complex structure which includes forget ($f^{<t>}$), update ($i^{<t>}$) and output ($o^{<t>}$) parameters and an additional output to the activations, the memory cell ($c^{<t>}$). Each gate performs the following equations:

\begin{equation}
\tilde c^{<t>} = \tanh(W_c \cdot [a^{<t-1>},x^{<t>}] + b_c)
\end{equation} 

\begin{equation}
f^{<t>} = \mbox{sigmoid}(W_u \cdot [a^{<t-1>},x^{<t>}] + b_u)
\end{equation}

\begin{equation}
i^{<t>} = \mbox{sigmoid}(W_f \cdot [a^{<t-1>},x^{<t>}] + b_f)
\end{equation}

\begin{equation}
o^{<t>} = \mbox{sigmoid}(W_o \cdot [a^{<t-1>},x^{<t>}] + b_o)
\end{equation} 
\begin{equation}
c^{<t>} = f^{<t>} * \tilde c^{<t>} + i^{<t>} * c^{<t-1>}
\end{equation} 
\begin{equation}
a^{<t>} = o^{<t>} * \tanh( c^{<t>})
\end{equation} 

\noindent
where $W_c$, $W_u$, $W_f$, and  $W_o$ stand for the parameter matrices; $b_c$, $b_u$, $b_f$, and $b_o$ for the biases; and $a^{<t>}$, $x^{<t>}$, and $c^{<t>}$ for the activation, input, and memory cell values, respectively, on the time stamp $t$.

\begin{figure}[h!]
	\centering
	\includegraphics[width=0.45\textwidth]{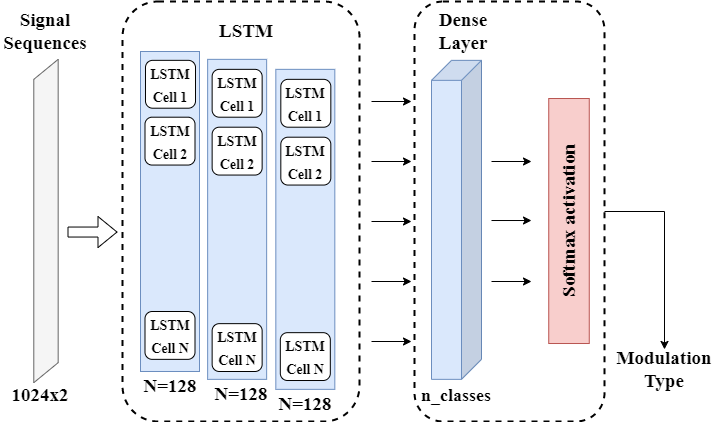}
	\caption{Proposed framework based on three stacked LSTM layers and a fully connected layer for signal classification.}
	\label{fig:framework}
\end{figure}

Following the idea that LSTM networks are ideal for processing sequences, our proposal consists of an LSTM Recurrent Neural Network over the signal's time series. The simplified network is made up of only three stacked LSTM layers of 128 cells each before the output layer. The first two LSTM layers return all sequence values (1024x128), and the final layer only returns a single value for the whole sequence and for each memory cell (1x128). Finally, the classification layer is a dense layer with a softmax activation function that has \textit{n\_classes} output neurons, a parameter that changes depending on the dataset, see Figure \ref{fig:framework}. Additionally, its implementation in TensorFlow (Version 2.4.1) consists of 330240 parameters in addition to the classification layer parameters, which can be 1032, 2967 or 3096 for 8, 23 and 24 classes respectively.
Regarding hyperparameters, we considered a batch size of 256 samples and the Adam optimizer with a cyclical learning rate to prevent convergence to local optima with values oscillating between \(1 \cdot 10^{-7}\) and \(1 \cdot 10^{-3}\). Finally, the network was trained from scratch in 300 epochs.
 This framework was designed to be trained, validated, and tested with the proposed datasets. To evaluate the performance of this neural network, three main metrics are used: 

\begin{itemize}
	\item Average accuracy of all classes with respect to the SNR.
	\item Minimum SNR (sensitivity) at which 90\% classification accuracy is achieved.
	\item Confusion matrices with respect to the SNR.
\end{itemize}
\section{Results}
\begin{figure}[H]
	\centering
	\includegraphics[width=0.45\textwidth]{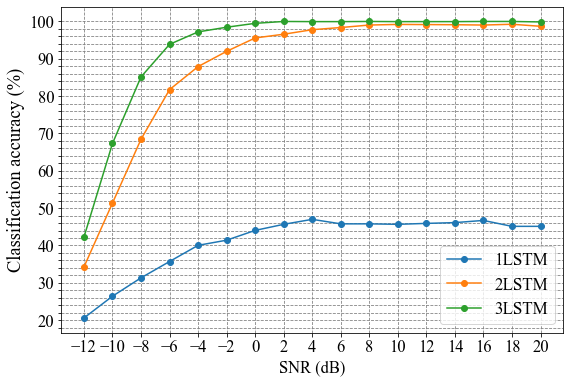}
	\caption{Comparison of the performance of the network for 1, 2 and 3 stacked LSTM layers using the 8-class radar dataset \cite{b44}.}
	\label{fig:layers}
\end{figure}

This section presents the performance of the proposed framework as a function of different parameters, for two radar datasets and one communications dataset. First, the performance of the network is evaluated depending on the simplicity of the structure by modifying the number of LSTM layers. To do so, training was performed on the radar dataset of eight modulations, considering that it is the simplest case with fewer number of classes. Figure \ref{fig:layers} shows that there is a significant gap between the performance with 1 layer and 2 and 3 LSTM layers. Since the best results are obtained with the 3-layer LSTM network, this structure will be further used to evaluate the performance of the radar and communications datasets. 

\subsection{Radar Signals: 8-class dataset}
\begin{figure}[h!]
	\centering
	\includegraphics[width=0.45\textwidth]{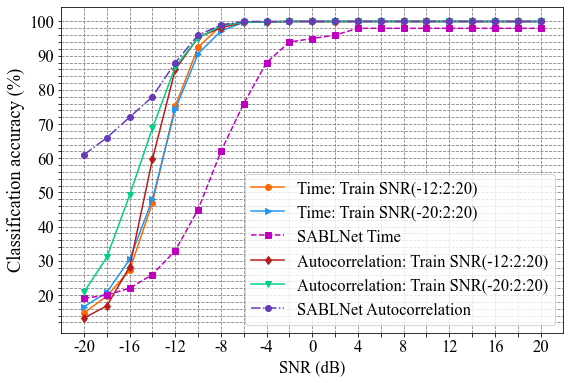}
	\caption{Average classification accuracy compared to SABLNet, studying the impact of the domain and the SNR range of the input data compare with the 8-class dataset \cite{b44}.}
	\label{fig:8class_avg}
\end{figure}

\begin{figure}[h!]
	\centering
	\includegraphics[width=0.45\textwidth]{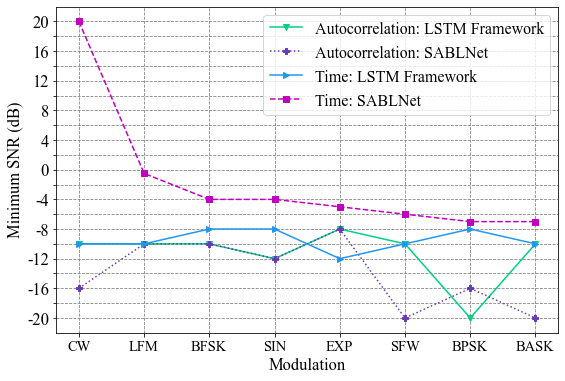}
	\caption{Comparison of the sensitivity for each signal modulation of our framework and SABLNet for sequences in time and autocorrelation domain  using the 8-class radar dataset \cite{b44}.}
	\label{fig:8class_sens}
\end{figure}

The performance of our framework is compared with the results obtained by the authors of the 8-class dataset and the SABLNet framework\cite{b44}. Their proposed structure is of higher complexity with CNN, Bi-LSTM and attention layers. \\
This section shows the results of two studies. First, we considered the impact of the signal domain, whether the network achieves better results with signals in time or in the autocorrelation domain. Our results do not show a significant improvement for sequences in the autocorrelation domain compared to raw time sequences, suggesting that our network learns from the original signal information, contrary to what the authors stated in \cite{b44}; see Figures \ref{fig:8class_avg} and \ref{fig:8class_sens}. Although the average classification accuracy does not improve for lower SNRs compared to SABLNet results, the minimum SNR required to obtain a 90\% classification accuracy remains at -10 dB for our network and SABLNet.
\\
Second, we reviewed the impact of the SNR range input data, when varying the SNR of the training samples from -12:2:20 dB to -20:2:20 dB and testing with the entire range of -20 to 20 dB, see Figure \ref{fig:8class_avg}. Under these circumstances, our network seemseeeee to be robust and insensitive while maintaining the classification accuracy for high-SNR scenarios.

\subsection{Radar Signals: DeepRadar2022}

In this section, we used our synthetic dataset DeepRadar2022, a more complex dataset with 23 signal types (see Table \ref{tab1}). 

\begin{figure}[H]
	\centering
	\includegraphics[width=0.45\textwidth]{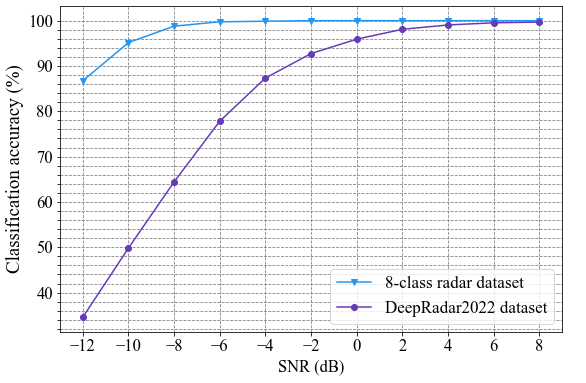}
	\caption{Comparison of the average classification accuracy for sequences in the time domain using DeepRadar2022 and the 8-class radar dataset \cite{b44}. (\textit{The curves remain still after 8 dB, therefore the SNR axis has been cropped})}
	\label{fig:23_overall}
\end{figure}
\begin{figure}[h!]
	\centering
	\includegraphics[width=0.45\textwidth]{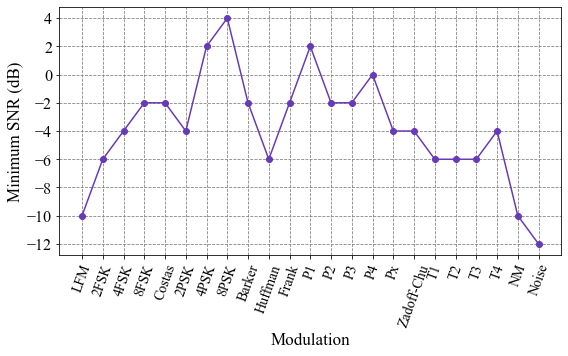}
	\caption{Sensitivity at 90\% classification accuracy for each signal modulation using the DeepRadar2022 dataset.}
	\label{fig:23_sensitivity}
\end{figure}
\begin{figure}[h!]
	\centering
	\includegraphics[width=0.45\textwidth]{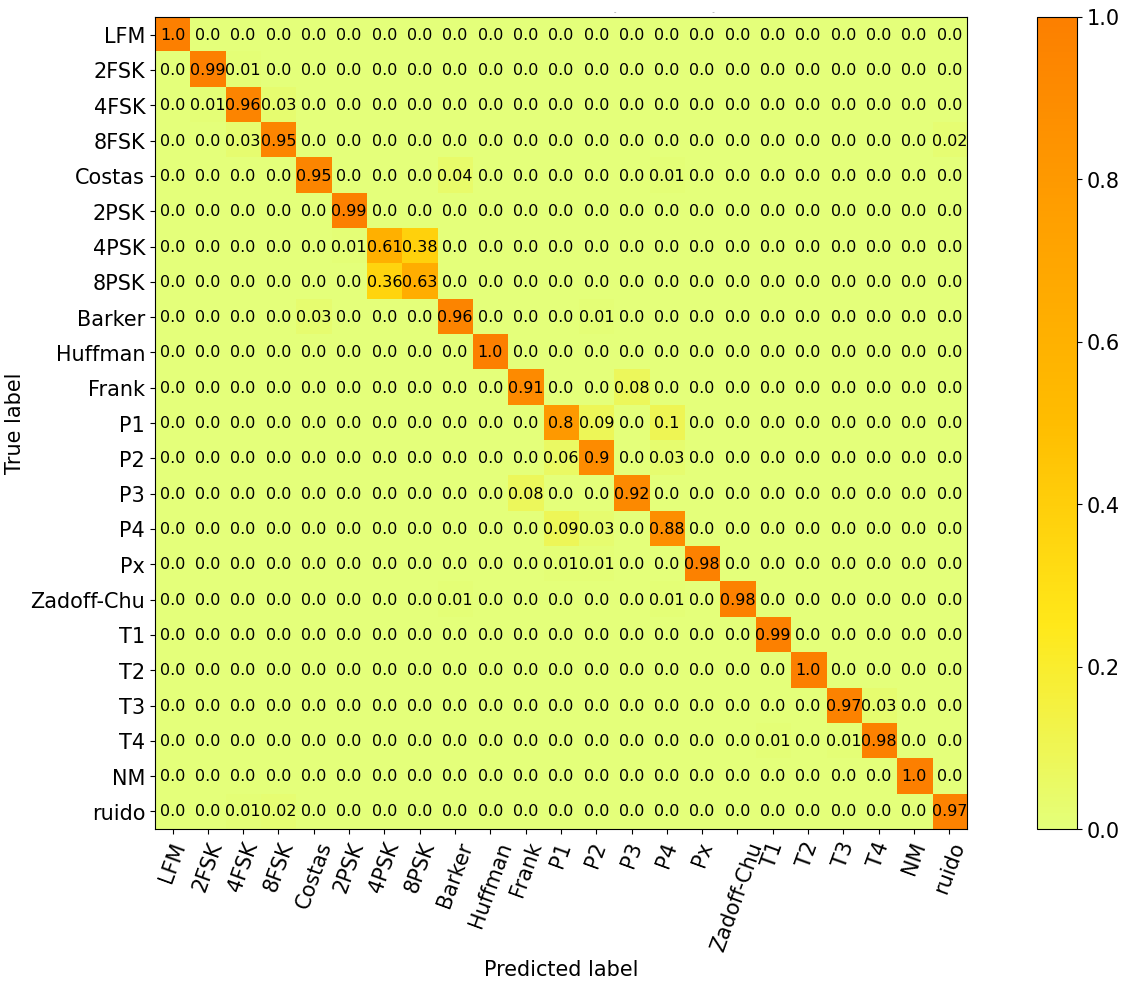}
	\caption{Confusion matrix for the DeepRadar2022 dataset at SNR = -2 dB that shows the misclassification of high-order phase modulations.}
	\label{fig:23_cm}
\end{figure}

As a result of the increase in the number of classes, a poorer sensitivity was obtained, from -10 dB with the 8-class dataset to -2dB with DeepRadar2022, see Figure \ref{fig:23_overall}.
Additionally, the sensitivity graph in Figure \ref{fig:23_sensitivity} has certain peaks for high-order continuous wave PSKs (4PSK and 8PSK). This idea is also outlined in the confusion matrix in Figure \ref{fig:23_cm}, where 4PSK and 8 PSK are misclassified as the other. This limitation could be improved by classifying all of them as PSK and discerning the order afterwards. Similarly, some signals with phase modulations using Frank code and codes P1 to P4 are incorreclty classified as the other.
\subsection{Communications Signals}
\begin{figure}[h!]
	\centering
	\includegraphics[width=0.45\textwidth]{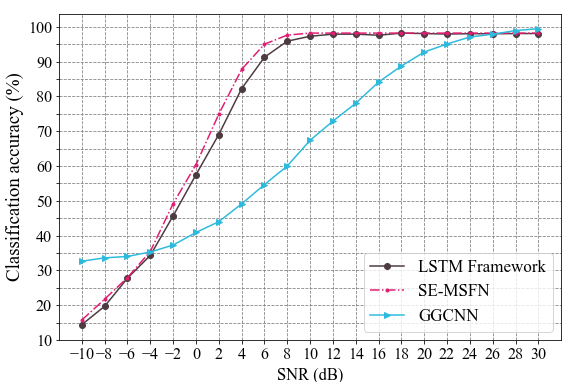}
	\caption{Average classification accuracy of our framework compared to  GGCNN \cite{b40} and SE-MSFN \cite{b41} using the RadioML 2018.01A dataset.
	\label{fig:comp_radioml}}
\end{figure}

Our network was trained and tested with the RadioML 2018.01A dataset of 24 modulation types. In their work, a network with residual stacks was proposed \cite{b8}. Similarly, we compared the average classification accuracy with the results obtained by the authors of GGCNN \cite{b40} and SE-MSFN \cite{b41}, see Figure \ref{fig:comp_radioml}. The results showed that an SNR of 6 dB is needed to obtain 90\% of classification accuracy. This value is high because of the difficulty in classifying high-order PSK and QAM and discerning between AM-DSB and AM-SSB with carrier (WC) and suppressed carrier (SC). Compared to the other proposals, the performance of our framework is superior but similar to that of SE-MSFN. However, in terms of complexity, our framework is simpler, avoiding a feature-extraction step and within fewer number of layers. 

\begin{figure}[h!]
	\centering
	\includegraphics[width=0.45\textwidth]{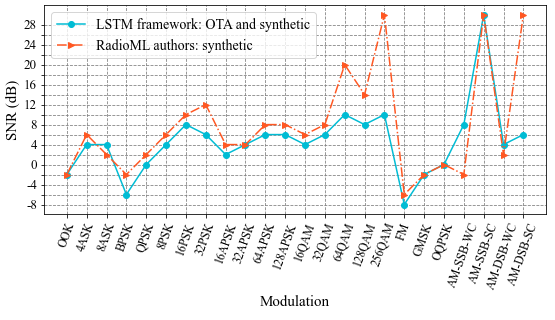}
	\caption{Comparison of the sensitivity at 90\% classification accuracy of our framework and the authors of the RadioML 2018.01A dataset \cite{b8}.
	\label{fig:sens_radioml}}
\end{figure}

Furthermore, we compared the sensitivity of each signal with the authors of RadioML 2018.01A, as they provided results related to the classification accuracy of each modulation. However, since we do not have the data set broken down by type of signal (synthetic or OTA capture) as in the original paper \cite{b8}, a fair comparison of the results is not possible. Despite that, by analyzing the results of the authors when classifying only the synthetic signals, some of the signals (64QAM, AM-DSB-SC, AM-SSB-SC and 256QAM) do not reach a 90\% accuracy at the highest evaluated SNR (20 dB). In contrast, our experiments show that all signals except AM-SSB-SC reach a 90\% classification accuracy for a SNR lower than 10 dB, see Figure \ref{fig:sens_radioml}. Therefore, the sensitivity of our framework seems to be more robust for most of the signals. 

\section{Conclusion}

 We have developed an LSTM-based architecture for the classification of
communications and radar signals. Our experiments show that
this framework achieves state-of-the-art performance for both types of signals.

These LSTM-based classifiers offer higher sensitivity and robustness
compared to current deep learning-based AMCs
with a simpler structure and, therefore, a reduced number of trainable
parameters.


\begin{thebibliography}{00}

\bibitem{b0} Zhechen Zhu and Asoke K. Nandi. 2015. Automatic Modulation Classification: Principles, Algorithms and Applications (1st. ed.). Wiley Publishing. 2015.

\bibitem{b1} O. A. Dobre, A. Abdi, Y. Bar-Ness, and W. Su, ``Survey of automatic modulation classification techniques: Classical approaches and new trends,'' in IET. 1. 137 - 156. 

\bibitem{b3} T. Wan, K. Jiang, Y. Tang, Y. Xiong and B. Tang, ``Automatic LPI Radar Signal Sensing Method Using Visibility Graphs," in IEEE Access, vol. 8, pp. 159650-159660, 2020.

\bibitem{b4} W. Gongming, C. Shiwen, H. Xueruobai and Y. Junjian, ``Radar Emitter Sorting and Recognition Based on Time-frequency Image Union Feature," 2019 IEEE 4th International Conference on Signal and Image Processing (ICSIP), Wuxi, China, 2019, pp. 165-170.

\bibitem{b5} J. Lunden and V. Koivunen, 
 ``Automatic Radar Waveform Recognition," in IEEE Journal of Selected Topics in Signal Processing, vol. 1, no. 1, pp. 124-136, June 2007.

\bibitem{b6} S. Rajendran, W. Meert, D. Giustiniano, V. Lenders and S. Pollin, ``Deep Learning Models for Wireless Signal Classification With Distributed Low-Cost Spectrum Sensors," in IEEE Transactions on Cognitive Communications and Networking, vol. 4, no. 3, pp. 433-445, Sept. 2018.

\bibitem{b44} S. Wei, Q. Qu, et al., ``Self-attention Bi-LSTM Networks for radar signal modulation recognition,'' in IEEE Transactions on Microwave Theory and Techniques, 69, 11 , pp 5160–-5172, 2021.

\bibitem{b7} T. Erpek, T. O'shea, Y. Sagduyu, Y. Shi and T. Charles Clancy,  ``Deep learning for wireless communications,'' in Development and Analysis of Deep Learning Architectures Studies in Computational Intelligence, pp 223--266, 2019.

\bibitem{b8} T. J. O’Shea, T. Roy and T. C. Clancy, ``Over-the-Air Deep Learning Based Radio Signal Classification," in IEEE Journal of Selected Topics in Signal Processing, vol. 12, no. 1, pp. 168-179, Feb. 2018.

\bibitem{b9}D. Li, R. Yang, X. Li and S. Zhu, ``Radar Signal Modulation Recognition Based on Deep Joint Learning," in IEEE Access, vol. 8, pp. 48515-48528, 2020.

\bibitem{b10} Z. Qu, W. Wang, C. Hou and C. Hou, ``Radar Signal Intra-Pulse Modulation Recognition Based on Convolutional Denoising Autoencoder and Deep Convolutional Neural Network," in IEEE Access, vol. 7, pp. 112339-112347, 2019.

\bibitem{b11} T. O'shea , J. Corgan, and T. Charles Clancy,  ``Convolutional radio modulation recognition networks,'' in Engineering Applications of Neural Networks Communications in Computer and Information Science, pp 213--226, 2016. O'Shea, Tim \& Corgan, Johnathan \& Clancy, T.. (2016). Convolutional Radio Modulation Recognition Networks. 

\bibitem{b27} W. Yongshi, G. Jie, L. Hao, L. Li, W. Zhigang and W. Houjun, ``CNN-based modulation classification in the complicated communication channel," 2017 13th IEEE International Conference on Electronic Measurement \& Instruments (ICEMI), Yangzhou, China, 2017, pp. 512-516.

\bibitem{b28} S. Peng et al., ``Modulation Classification Based on Signal Constellation Diagrams and Deep Learning," in IEEE Transactions on Neural Networks and Learning Systems, vol. 30, no. 3, pp. 718-727, March 2019.

\bibitem{b29} K. Yashashwi, A. Sethi and P. Chaporkar, ``A Learnable Distortion Correction Module for Modulation Recognition," ssh  in IEEE Wireless Communications Letters, vol. 8, no. 1, pp. 77-80, Feb. 2019.

\bibitem{b30}R. Li, L. Li, S. Yang and S. Li, ``Robust Automated VHF Modulation Recognition Based on Deep Convolutional Neural Networks," in IEEE Communications Letters, vol. 22, no. 5, pp. 946-949, May 2018.

\bibitem{b31} Z. Liu, L. Li, H. Xu and H. Li, ``A method for recognition and classification for hybrid signals based on Deep Convolutional Neural Network," 2018 International Conference on Electronics Technology (ICET), Chengdu, China, 2018, pp. 325-330.

\bibitem{b32} D. Wang, M. Zhang, J. Li et al.,  ``Intelligent constellation
diagram analyzer using convolutional neural network-based
deep learning,'' in Optics Express, 25, 15, pp 17150–-17166, 2017.

\bibitem{b33} S. Peng, H. Jiang, H. Wang, H. Alwageed and Y. -D. Yao, ``Modulation classification using convolutional Neural Network based deep learning model," 2017 26th Wireless and Optical Communication Conference (WOCC), Newark, NJ, USA, 2017, pp. 1-5.

\bibitem{b34} H.Wu, Q.Wang, L. Zhou et al.,  ``VHF radio signal modulation
classification based on convolution neural networks,'' in Matec
Web of Conferences, 246, 2018.

\bibitem{b35} Z. Liu, L. Li, H. Xu, and H. Li,  ``A method for recognition and
classification for hybrid signals based on deep convolutional
neural network,'' in Proceedings of the 2018 International Conference
on Electronics Technology (ICET), pp 325-–330, 2018.
\bibitem{b40}P. Ghasemzadeh, M. Hempel, H. Wang and H. Sharif, ``GGCNN: An Efficiency-Maximizing Gated Graph Convolutional Neural Network Architecture for Automatic Modulation Identification," {\em IEEE Transactions On Wireless Communications}. pp. 1-1 (2023)
\bibitem{b41} X. Wu, S. Wei and Y. Zhou, ``Deep Multi-Scale Representation Learning with Attention for Automatic Modulation Classification,"{\em 2022 International Joint Conference on Neural Networks (IJCNN)}, Padua, Italy, 2022, pp. 1-8.

\bibitem{b15} C. Wang, J. Wang, and X. Zhang,  ``Automatic radar waveform recognition
based on time-frequency analysis and convolutional neural network,'' in 2017 IEEE In-
ternational Conference on Acoustics, Speech and Signal Processing (ICASSP), 2017.

\bibitem{b23} J. Wan, X. Yu and Q. Guo,  ``LPI radar waveform recognition based
on CNN and TPOT,'' in  Symmetry, 11, 5, 2019.

\bibitem{b24} Z. Qu, C. Hou, C. Hou and W. Wang, ``Radar Signal Intra-Pulse Modulation Recognition Based on Convolutional Neural Network and Deep Q-Learning Network,'' in IEEE Access, vol. 8, pp. 49125-49136, 2020.

\bibitem{b26} Z. Zhou, G. Huang and H. Chen,  ``Automatic radar waveform
recognition based on deep convolutional denoising auto-encoders,'' in Circuits Systems and Signal Processing, 37, 9, pp 4034--4048, 2018.

\bibitem{b25} Z. Liu, X. Mao, Z. Deng, et al.,  ``Radar signal waveform recognition
based on convolutional denoising autoencoder,'' in Proc. International
Conference on Communications, pp 752--761, 2018.

\bibitem{b22} Q. Wang, P. Du, J. Yang, et al.,  ``Transferred deep learning based
waveform recognition for cognitive passive radar,'' in  Signal Processing, pp 259–-267, 2019.

\bibitem{b21} S. Wei, Q. Qu, H. Su, et al.,  ``Intra-pulse modulation radar signal recognition
based on CLDN network,'' in  IET Radar, Sonar and Navigation, 14, 6, pp 803–-810, 2020.

\bibitem{b16} X. Li, Z. Liu, and Z. Huang,  ``Attention-based radar PRI modulation recognition with Recurrent Neural Networks,'' in IEEE Access, 8, pp 57426-–57436, 2020.

\bibitem{b17} Z. Liu and P. Yu,  ``Classification, denoising, and deinterleaving of pulse streams with recurrent neural networks,'' in IEEE Transactions on Aerospace and Electronic Systems, 4, 55, pp 1624–-1639, 2019.

\bibitem{b18} X. Li, Z. Liu, Z. Huang and W. Liu,  ``Radar emitter classification with attention-based Multi-rnns,'' in  IEEE Communications Letters, 9, 24, pp 2000–-2004, 2020.

\bibitem{bb4} S. Hochreiter, J. Schmidhuber, ``Long short-term memory,'' in Neural Computation, vol. 9, no. 8, pp. 1735-1780, 1997.

\bibitem{b20} L. Smith,  ``Cyclical learning rates for training neural networks,'' in  2007 IEEE Winter Conference on Applications of Computer Vision (WAVC), 9, 24, pp 2000–-2004, 2017.

\end{thebibliography}
\end{document}